\documentclass[fleqn]{2017SCGE}
\makeatletter
\renewcommand*{\@fnsymbol}[1]{\ensuremath{\ifcase#1\or \dagger \else\@ctrerr\fi}}
\makeatother

\begin{document}

\ensubject{subject}

\ArticleType{Article}
\SpecialTopic{SPECIAL TOPIC: }
\Year{2021}
\Month{XXX.}
\Vol{X}
\No{X}
\DOI{XX.XXXX/XXXXXX-XXX-XXXX-X}
\ArtNo{000000}
\ReceiveDate{XXX. XX, 2021}
\AcceptDate{XXX. XX, 2021}


\title{A giant central red disk galaxy at redshift $z=0.76$: challenge to theories of galaxy formation}{A giant central red disk galaxy: challenge to galaxy formation theories}

\author[1]{Kun XU}{}
\author[1]{ChengZe LIU}{czliu@sjtu.edu.cn}
\author[1,2]{YiPeng JING}{ypjing@sjtu.edu.cn}
\author[3]{Marcin SAWICKI$^\dagger$}{}\thanks{Canada Research Chair}
\author[4]{Stephen GWYN}{}

\AuthorMark{Xu K}

\AuthorCitation{Xu K, Liu C Z, Jing Y P, et al}

\address[1]{Department of Astronomy, School of Physics and Astronomy, Shanghai Jiao Tong University, Shanghai {\rm 200240}, China}
\address[2]{Tsung-Dao Lee Institute, and Shanghai Key Laboratory for Particle Physics and Cosmology, Shanghai Jiao Tong University, \\Shanghai {\rm 200240}, China}
\address[3]{Department of Astronomy \& Physics and the Institute for Computational Astrophysics, Saint Mary's University, 923 Robie Street, \\Halifax {\rm Nova Scotia B3H 3C 3}, Canada}
\address[4]{NRC-Herzberg Astronomy and Astrophysics, National Research Council of Canada, 5071 West Saanich Rd., Victoria {\rm BC V9E 2E7}, Canada}


\abstract{We report a giant red central disk galaxy in the XMM-LSS north region. The region is covered with a rich variety of multi-band photometric and spectroscopic observations. Using the photometric data of the Canada-France-Hawaii Telescope Legacy Survey (CFHTLS) and spectroscopic observation of the Baryon Oscillation Spectroscopic Survey (BOSS), we find that the galaxy has a stellar mass of $\sim10^{11.6}$ M$_{\odot}$. The galaxy has a red color and has an old stellar population, and thus its star formation has stopped. With the photometric image data of Hyper Suprime-Cam (HSC) Subaru Strategic Program, we demonstrate that its luminosity profile is perfectly described by a S\'ersic form with n$=1.22$ indicating disk morphology. We also analyze its environment based on the VIMOS Public Extragalactic Redshift Survey (VIPERS) photometric catalog, and find that its close neighbors are all less massive, indicating that our observed galaxy is sitting at the center of its host halo. Existence of the giant red central disk galaxy seriously challenges the current standard paradigm of galaxy formation, as there is no known physical mechanism to explain the quenching of its star formation. This conclusion is supported by state-of-the-art hydrodynamical simulations of galaxy formation.}

\maketitle


\begin{multicols}{2}
\section{Introduction}

Galaxies are usually classified into ellipticals and spirals according to their morphology. Elliptical galaxies, by definition, have regular elliptical shapes of stellar distribution. They have very little ongoing star formation, and thus they are red in color and old in stellar age. In contrast, spiral galaxies have disk shapes of stellar distribution, 
\Authorfootnote
\noindent 
though a significant fraction of them possess bulges at the centers. There is usually ongoing star formation in spiral galaxies, so they look blue in color and have young stellar populations \cite{ref1}. 

These properties of galaxies can be well understood in the standard paradigm of hierarchical structure formation~\cite{ref2,ref3,ref4}. In the theory, cosmic structures grow by gravitational instability from the initial tiny quantum fluctuations generated during the inflationary epoch. Dark Matter (DM) halos, which are defined as the dark matter objects with density about $200$ times of the mean density of the Universe, are formed through accreting surrounding smaller DM halos and diffuse matter including DM and gas. The gas, when accreted into a halo, is heated by shocks \cite{shock}. The hot gas cools by radiating its energy, and the cold gas spirals into the halo center to form a disk (spiral) galaxy. With   
the growth of a halo, surrounding galaxies can also be accreted into the halo and become its satellites. Because of dynamical friction, some satellite galaxies, especially relatively massive ones, spiral into the center and coalesce with the centra galaxy~\cite{ref5,ref6}. A merger of a central disk galaxy with a satellite of comparable mass (say, a mass ratio $\geq 0.2$) may significantly change its internal structure, either producing  an elliptical galaxy or a disk galaxy with a significant bulge\cite{ref7,ref8,ref9}. Based on the observed relation between black hole mass and bulge mass \cite{ref10}, we expect there are supermassive black holes (SMBHs) within elliptical galaxies or bulges. The strong energy output and/or material outflow produced by an SMBH can heat and/or blow out the surrounding cold gas, thus suppressing the reservoir from which stars form and making the host galaxy look red. Current hydrodynamical simulations of galaxies that have incorporated these (and other) physical processes have successfully reproduced a wide range of observed properties of galaxies including the stellar mass, morphology, color, and stellar age~\cite{ref11,ref12}.
\begin{figure}[H]
\centering
\includegraphics[scale=0.28]{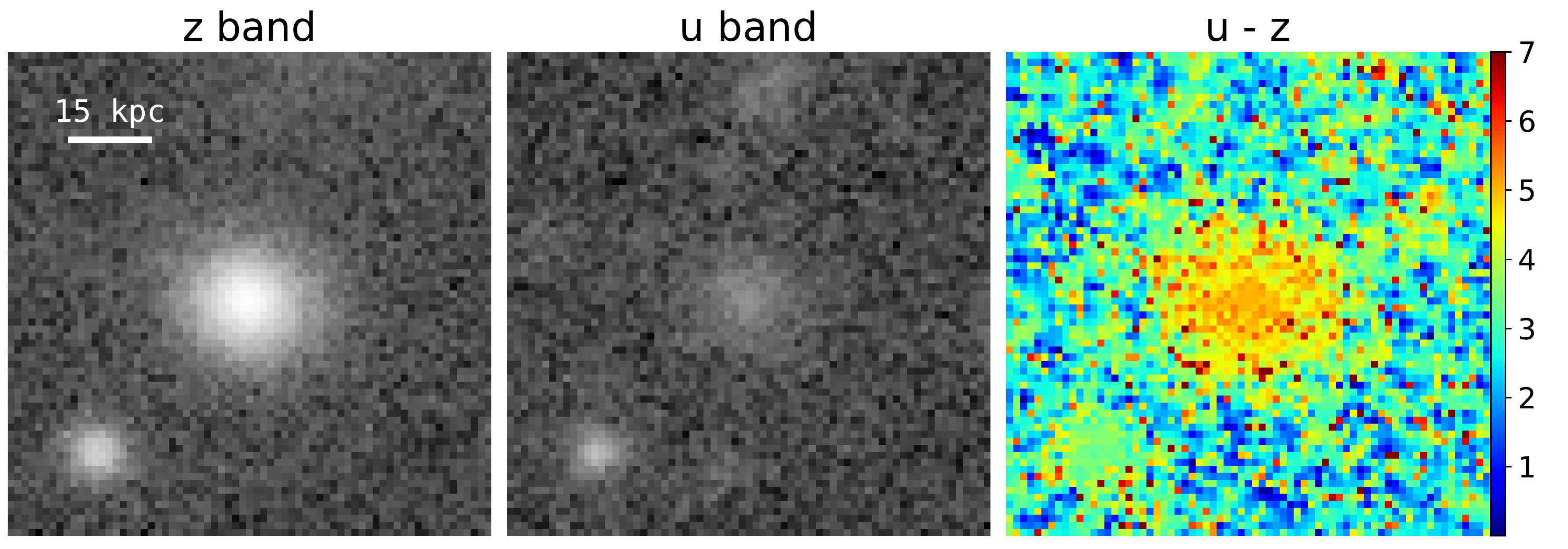}
\caption{HSC $z$-band image (left panel), CLAUDS $u^{*}$-band image(middle panel) and $u^{*}-z$ color magnitude map (right panel) of the galaxy. Physical length of 15 $kpc$ corresponds to 1.977 $arcsec$ and the FWHM of the seeing in the $z$ and $u^*$ images is 0.822 $arcsec$ and 1.046 $arcsec$.}\label{fig:f1}
\end{figure}
Massive galaxies with stellar mass larger than  $10^{11.0}M_{\odot}$  are expected to form in DM halos of group and cluster mass~\cite{ref13}. Because of frequent mergers happening in such rich environments, one expects that most such massive galaxies are red ellipticals with significant amounts of bulge. Only in a rare case with little merger happened in the past, because no bulge has formed, thus no SMBH has formed. The central galaxy may keep forming stars, thus becoming a blue disk galaxy. In any case, one would not expect to form a central red disk galaxy in a massive DM halo, as there is no mechanism to stop gas cooling and star formation if no bulge (thus no SMBH) has formed.
\begin{table}[H]
    \centering
    \caption{Physical properties of SDSS 022248.95-045914.9.}
    \begin{tabular}{cc}
         \toprule   
         Properties & value \\
         \midrule   
         RA ($deg$) & $35.703888$ \\
         DEC ($deg$) & $-4.987444$ \\
         Redshift & $0.76$ \\
         rest-frame $NUV-r$ ($mag$) & $5.3$ \\
         sSFR ($yr^{-1}$) & $10^{-11.38}$ \\
         $M_*$ ($M_{\odot}$) (SED) & $10^{11.64 \pm 0.10}$ \\
         $M_*$ ($M_{\odot}$) (PCA\textsuperscript{a}) & $10^{11.67 \pm 0.17}$ \\
         S\'ersic index & $1.22 \pm 0.06$ \\
         axis ratio $b/a$ & $0.85$ \\
         half-light radius $R_e$ (kpc) & $8.81$ \\ 
         $NUV$ ($mag$)& $26.9$ \\
         $u$ ($mag$) & $24.7$ \\
         $g$ ($mag$) & $22.8$ \\
         $r$ ($mag$) & $21.3$ \\
         $i$ ($mag$) & $20.0$ \\
         $z$ ($mag$) & $19.5$ \\
         $K_s$ ($mag$) & $18.4$\\
         \bottomrule  
         \multicolumn{2}{l}{\textsuperscript{a}\footnotesize{From Wisconsin group \cite{ref16}}}
    \end{tabular}
    \label{tab:t1}
\end{table}
\section{Properties of the galaxy}

\begin{figure}[H]
\centering
\includegraphics[scale=0.34]{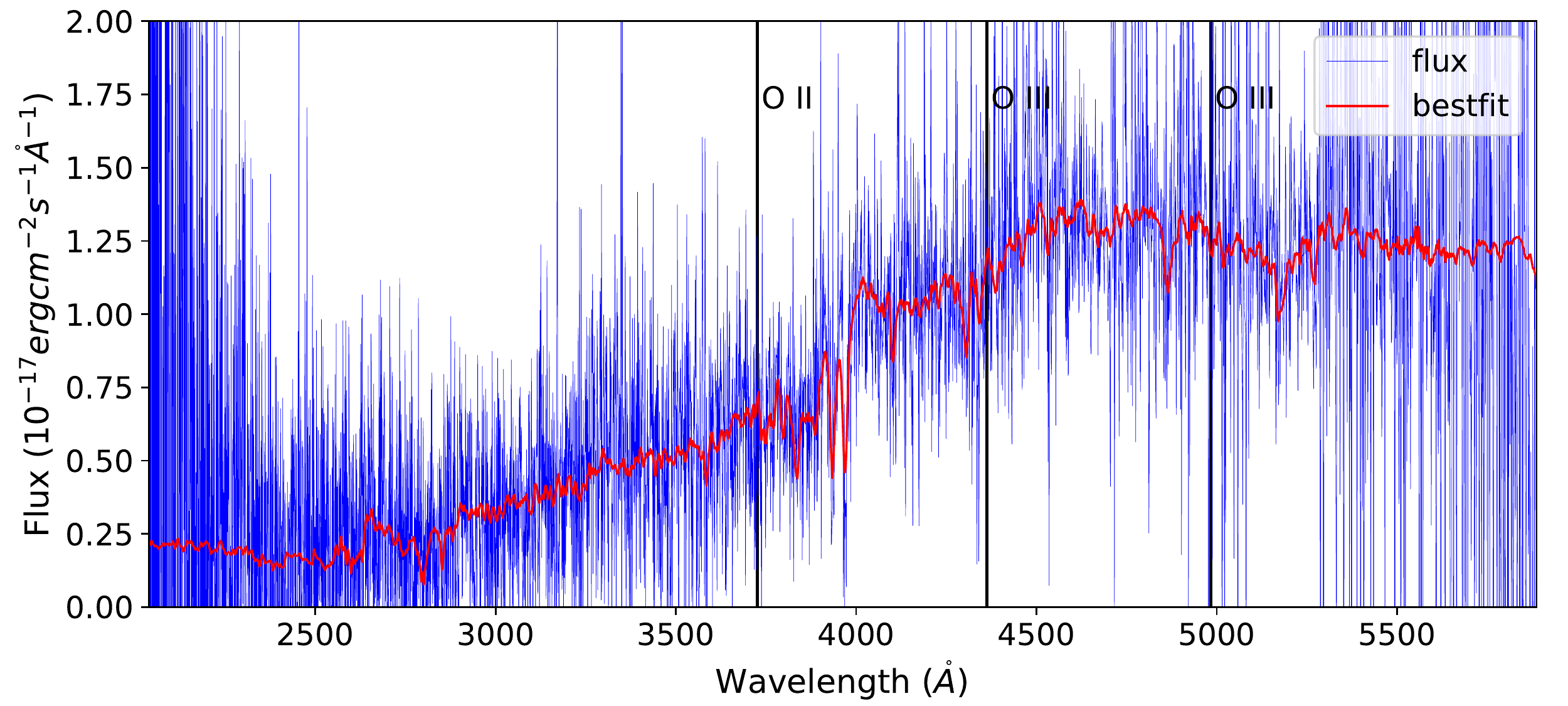}
\caption{Blue line shows the original spectrum from BOSS and the red line is the best fitting model. Typical emission lines are shown via vertical black lines. Wavelength is in rest-frame.}\label{fig:f2}
\end{figure}

In this paper, we report a massive red central disk galaxy SDSS J022248.95-045914.9 with stellar mass reaching $10^{11.6}M_{\odot}$. The details of the galaxy are shown in Table \ref{tab:t1}. The $z$-band image from Deep field of HyperSuprime-Cam Subaru Strategic Program (HSC-SSP) \cite{ref18}, $u^{*}$-band image from CFHT Large Area U-band Deep Survey (CLAUDS) \cite{clauds}, and $u^{*}-z$ color map of the galaxy are shown in Figure \ref{fig:f1}. The galaxy was discovered during our study of physical properties of massive galaxies in the CMASS sample of the Baryon Oscillation Spectroscopic Survey (BOSS) survey \cite{ref14} in the XMM-LSS region \cite{ref33}. We perform Spectral Energy Distribution (SED) fitting on multi-band data of the galaxy from $FUV$ to $K_s$ collected by Moutard et al. \cite{ref15} using  \textsc{Le PHARE} \cite{le1,le2}. We use the stellar population synthesis models of BC03 \cite{bc03} with the Chabrier \cite{imf} Initial Mass Function (IMF) and a delayed star formation history. We get the stellar mass $M_{\ast}=10^{11.64}M_{\odot}$ and the specific star formation rate (sSFR) $10^{-11.38}\ {\rm yr^{-1}}$ and the rest-frame $NUV-r$ color is $5.3$ based on the SED fitting. As the BOSS survey has taken a spectrum for the galaxy, we show it in Figure \ref{fig:f2}. From the figure, we can easily see the best fitting spectrum is typical for an old stellar population: there is no visible emission line and the $D4000$ break is about $1.9$. The stellar mass released by the Wisconsin group \cite{ref16} using Principal Component Analysis method based on the spectrum is $10^{11.67\pm 0.17}M_{\odot}$, which is well consistent with the stellar mass obtained by our SED fitting.  From both the photometry and the spectrum, we can conclude that the galaxy is very massive and its star formation has been quenched. These are the typical features of a massive elliptical galaxy.

\begin{figure}[H]
\centering
\includegraphics[scale=0.3]{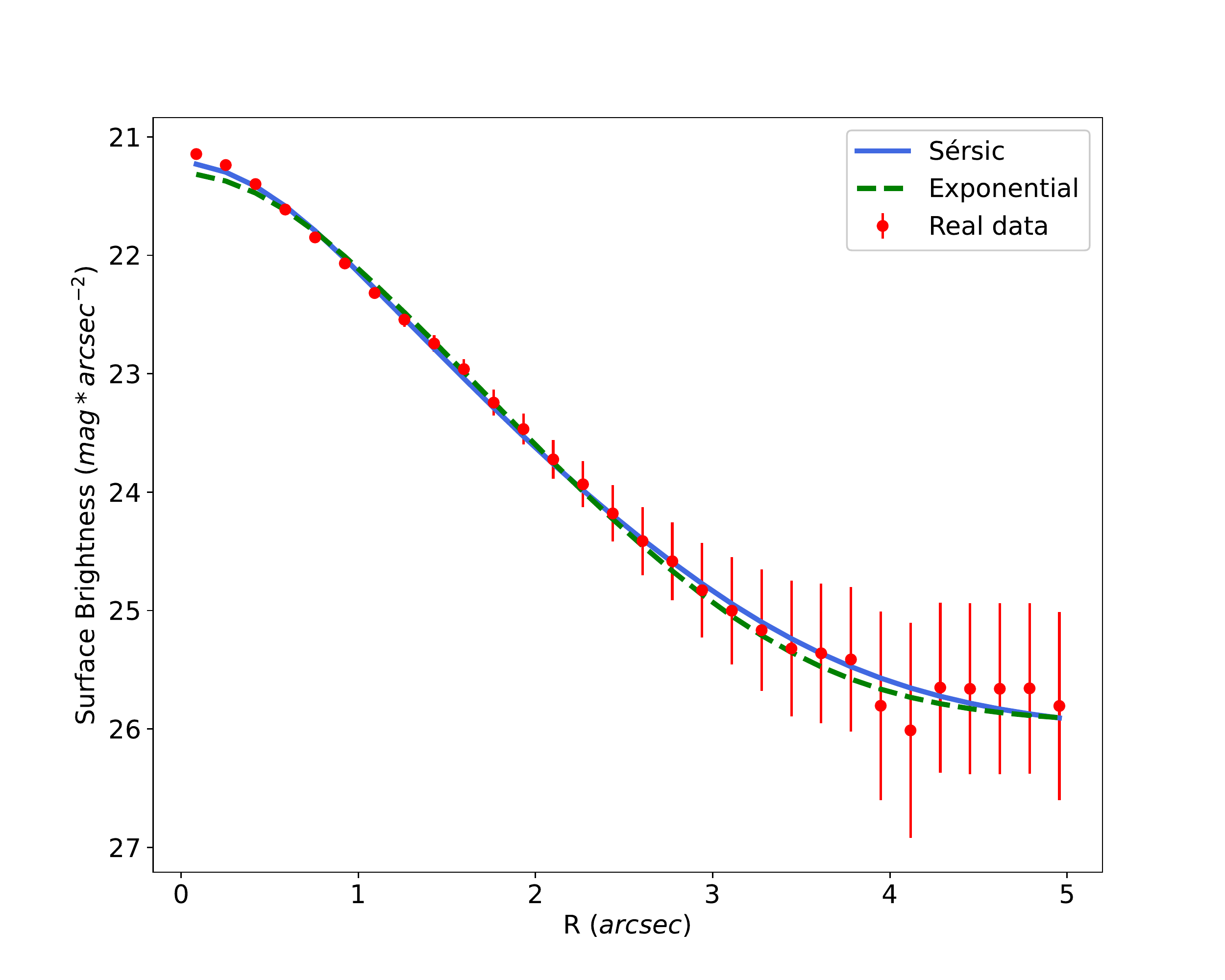}
\caption{Radial profile of the $z$ band image and models of the galaxy. Red dots with error bars are the data from HSC $z$ band. Blue solid line shows the profile of the best fitting S\'ersic model. Parameters from the best fitting are: Integrated magnitude: $z=19.64 \pm 0.01\ {\rm mag}$; effective radius: $R_e=1.035 \pm 0.020\ {\rm arcsec}$; Axis ratio: $b/a=0.850 \pm 0.010$; Position Angle: $PA= -78.46 \pm 3.87\ {\rm deg}$; S\'ersic index: $n= 1.22 \pm 0.06$. Green dot line shows the best fitting exponential function.}\label{fig:f3}. 
\end{figure}

However, our analysis of the morphology for the galaxy surprisingly shows that it is a disk galaxy. The radial luminosity distribution of a galaxy can be described by the S\'ersic form \cite{ref17}:
\begin{equation}
\label{eq:1}
I(R) = I_e\exp\bigg\{-b_n\bigg[\bigg(\frac{R}{R_e}\bigg)^{1/n}-1\bigg]\bigg\},
\end{equation}
where $b_n$ satisfies:
\begin{equation}
\label{eq:2}
\gamma(2n; b_n)=\frac{1}{2}\Gamma(2n).
\end{equation}
$I_e$ is the intensity at the half-light radius $R_e$, and $n$, called the S\'ersic index, describes the compactness of a luminosity profile. $\Gamma$ and $\gamma$ are respectively the Gamma function and lower incomplete Gamma function. Generally speaking, disk galaxies have exponential profiles with $n = 1$, and ellipticals follow the de Vaucouleurs profiles with $n = 4$. We take the $z$ band photometric image from the public data release $2$ of HyperSuprime-Cam Subaru Strategic Program (HSC-SSP) \cite{ref18}. We fit the image with the S\'ersic form in 2D using \textsc{galfit} \cite{ref19} with the point spread function (PSF) taken into account. The best-fitting S\'ersic profile of the galaxy is shown in Figure \ref{fig:f3}, which describes the observed radial luminosity profile very well. The S\'ersic index is $n = 1.22 \pm 0.06$, which indicates the galaxy is a disk galaxy. This is consistent with the visual appearance, which does not show any significant bulge component. To directly measure the bulge/disk ratio, we also performed two component fitting on the galaxy, using an exponential plus de Vaucouleurs profile. The fitting results do not indicate any bulge component, and we can set an upper limit that the bulge contains less than $1/15$ of the stellar light of the galaxy. In fact, the luminosity profile is indeed well described by the exponential profile with $n=1$ (Green dot line in Figure \ref{fig:f3}).

\begin{figure}[H]
\centering
\includegraphics[scale=0.45]{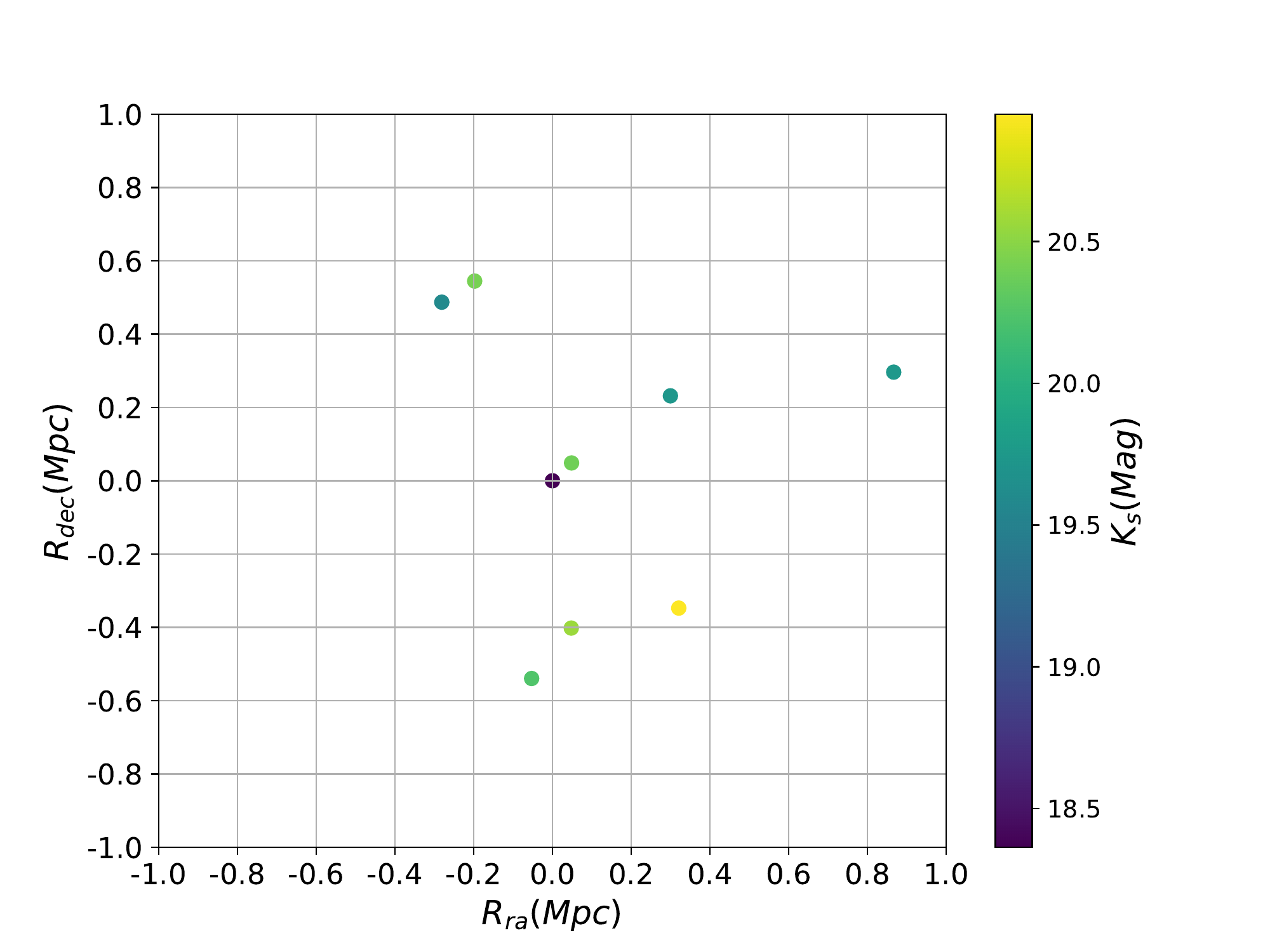}
\caption{Galaxies of $K_s<21$ $mag$ within $|\Delta z_{ph}| = 0.2$ in the sky area of a comoving radius $1\ {\rm Mpc}$ centered on the red disk galaxy. Colors of the dots reflect the $K_s$ band magnitude of the neighbor galaxies.}\label{fig:f4}
\end{figure}

As we will discuss below, the environment effects, such as the ram pressure and/or tidal stripping may help disk galaxies quench their star formation. It is therefore vital to check whether the galaxy is a central or satellite galaxy. We use the photometric catalog of the second public data release (PDR-2) of the VIMOS Public Extragalactic Redshift Survey (VIPERS) \cite{ref15}. This catalog contains all sources brighter than $22.5$ $mag$ in the $i$ band in the W1 field which covers the galaxy we study. The $K_s$ band magnitudes from the VIDEO observation \cite{ref20} are also available in the region around the galaxy. We consider a field of view of comoving radius $1\ {\rm Mpc}$ around the galaxy, which corresponds to $\theta=0.02081\  deg$ in the sky using the cosmological parameters from the Planck satellite \cite{ref21}. We use the photometric redshifts $z_{ph}$ to search for the neighbors. Since the error of  $z_{ph}$  is typically about $0.05$ for galaxies with $i<22.5$ $mag$ and $K_s<22.0$ $mag$ \cite{ref15}, we plot all the galaxies of $K_s\leq 21.0$ $mag$ within a cylinder of radius $\theta$ and length $\Delta z_{ph} =0.4$ centered on redshift $z_s=0.76$ of the red disk galaxy. As shown in Figure \ref{fig:f4}, there are $8$ neighbor galaxies, all of which are at least $1.2$ mag fainter than our red disk ($K_s=18.4$ $mag$), which confirms that the red disk galaxy is a central galaxy. Since the $K_s$ luminosity is approximately proportional to the stellar mass, we also find that the existence of $8$ satellites with the stellar mass larger than $1/10$ of the central red disk??s stellar mass is fully consistent with the observed conditional stellar mass function \cite{ref22}.
Thus, the galaxy we find is very likely a massive central red disk. But it is very challenging to explain its formation in current theories of galaxy formation. According to the halo mass and stellar mass relation \cite{ref13}, we expect the galaxy is at the center of the host halo of mass about $5\times 10^{14}M_{\odot}$. As described in the beginning, such a galaxy should find it hard to maintain its disk structure since frequent merging is inevitable with other spiraling-in galaxies in the rich environment (cluster of galaxies). Even in very rare of a galaxy with very little merging in its recent past, one would expect ongoing star formation activities unless there is enough feedback from the central SMBH. However, as we have not detected the bulge component at the center of the red disk, we would expect that the feedback from the SMBH is not an efficient way to quench the whole galaxy. This qualitative reasoning is verified by state-of-the-art hydrodynamical simulations that implement the physical processes we have described. We have searched for massive red disks ($NUV-r>4.5$ or sSFR $<10^{-11}\ {\rm yr^{-1}}$) of $M_{\ast} >10^{11.3}M_{\odot}$ and $n<1.5$ (in rest-frame $r$ band) in the Illustris-1 simulation \cite{ref23,ref24} at $z_s=0.5$, but we found a null result. To test the sensitivity to the details of simulations, we have also inspected the Illustris TNG100-1 \cite{ref25} simulation at $z_s = 0$ \footnote{Because the color of galaxies is redder and the merging is less frequent in a fixed time period at $z_s=0$ than $z_s=0.7$, we would expect to form more red disk galaxies at $z_s=0$ than at an earlier epoch.}, and no massive red disk is found either.
\section{Discussion and conclusion}

It is interesting to point out that a few studies of red spiral galaxies have appeared recently in the literature. Red disks with smaller mass ($\sim10^{10.0}M_{\odot}$) are found at intermediate local densities such as the infall regions of clusters~\cite{ref26,ref27}. The tidal force and/or gas ram pressure in such intermediate density environments could be strong enough to destroy the cold gas reservoir, but still too weak to affect the disk structures. Statistical studies of nearby massive red disks ($10^{10.5}\sim10^{11.0}M_{\odot}$) show larger bulge/total mass ratios than their blue counterparts~\cite{ref28,ref29}. Various quenching mechanisms have been theoretically proposed to quench a massive galaxy without AGN feedback such as halo quenching \cite{shock} and angular-momentum quenching \cite{angular}, which should be investigated further. In other studies, relations beyond the standard scaling rules are proposed to explain the existence of massive red disks. For example, some studies show that massive disks may exist in smaller halos comparing to other galaxies with the same stellar mass \cite{ref30}. Gas is exhausted by star formation in these galaxies and no quenching mechanism is needed to explain their red color. Moreover, deviations from the BH-bulge mass and the DM-baryon mass relations have also been proposed \cite{ref31,ref32}. Compared to the red galaxies found in the above studies, the galaxy reported in this paper has the highest redshift and, more importantly, the largest mass, making it much harder to explain its quenching.

In conclusion, we find a giant red central disk. The existence of this galaxy challenges current theories of galaxy formation, as there is no known physical mechanism to stop its star formation. Since there are some rare cases whereby mergers, which may cause very low S\'ersic indices, can not be distinguished from imaging data \cite{ref33}, we plan to confirm the galaxy's disk morphology with kinematic data from Integral Field Unit (IFU) observations. Meanwhile, we will search the whole HSC wide field for massive red disks and employ statistical analysis to check whether their properties deviate from standard scaling rules or not.    

\Acknowledgements{The work is supported by NSFC (11533006, 11621303, 11890691, 11933003, 11833005). We are grateful to Yong Shi, Xiaoyang Xia, Caina Hao, Yongzhong Qian, Yanmei Chen, and Yunchong Wang for their helpful discussions. This research uses data obtained through the Telescope Access Program (TAP), which has been funded by the National Astronomical Observatories, Chinese Academy of Sciences, and the Special Fund for Astronomy from the Ministry of Finance. This work is based on observations obtained with MegaPrime/MegaCam, a joint project of CFHT and CEA/DAPNIA, at the Canada-France-Hawaii Telescope (CFHT) which is operated by the National Research Council (NRC) of Canada, the Institut National des Science de l'Univers of the Centre National de la Recherche Scientifique (CNRS) of France, and the University of Hawaii. The work has made use of the public data of the Canada-France-Hawaii Telescope Legacy Survey, the Sloan Digital Sky Survey, the Hyper Suprime-Cam Subaru Strategic Program, and the VIMOS Public Extragalactic Redshift Survey.}

\InterestConflict{The authors declare that they have no conflict of interest.}



\end{multicols}
\end{document}